\documentclass[twocolumn,aps,prl,superscriptaddress,longbibliography]{revtex4-1}
\usepackage[colorlinks,bookmarks=false,citecolor=blue,linkcolor=red,urlcolor=blue]{hyperref}
\usepackage{verbatim}
\usepackage{color}
\usepackage{amsmath,amssymb,bm,braket,float,mathtools,subfigure}
\usepackage{graphicx,xfrac,appendix}
\usepackage[english]{babel}

\makeatletter

\usepackage{epstopdf}

\makeatother

\begin{document}
\title{Bulk-edge correspondence for the nonlinear eigenvalues problem of the Haldane model}
\author{Shujie Cheng}
\email{chengsj@zjnu.edu.cn}
\affiliation{Xingzhi College, Zhejiang Normal University, Lanxi 321100, China}
\affiliation{Department of Physics, Zhejiang Normal University, Jinhua 321004, China}
\author{Yonghua Jiang}
\email{yonghua$_$j82@zjnu.cn}
\affiliation{Xingzhi College, Zhejiang Normal University, Lanxi 321100, China}
\author{Gao Xianlong}
\email{gaoxl@zjnu.edu.cn}
\affiliation{Department of Physics, Zhejiang Normal University, Jinhua 321004, China}

\date{\today}

\begin{abstract}
Recently, there is an interest in studying the bulk-edge correspondence for nonlinear eigenvalues problems in a 
two-dimensional topological system with spin-orbit coupling. By introducing auxiliary eigenvalues, the nonlinear 
bulk-edge correspondence was established. In this paper, taking the Haldane model as an example, we address 
that such a correspondence will appear in two dimensional topological system without spin-orbit coupling. 
The resulting edge states are characterized by the Chern number of the auxiliary energy band. A full phase 
diagram containing topological nontrivial phase, topological trivial phase, and metallic phase is obtained.
Our work generalizes the study of the bulk-edge correspondence for nonlinear eigenvalue problems in two-dimensional system.
\end{abstract}

\maketitle

\section{introduction}
For decades, the novel nature of the topological phase of matter has sparked significant interest among researchers 
in the field of condense matter topology. Specifically, topological band theory plays critical role in revealing a variety 
of topological phases by integrating the energy band theory and the concept of topology \cite{1,2,3,4,5,6,7,8,9,10,11,12}. One of the most fascinating 
phenomena in topological systems is the bulk-edge correspondence (BEC), which showcases the appearance of edge states 
triggered by the bulk topology \cite{13,14}. One can employ the quantum transport to check this correspondence. The topological systems 
with non-zero quantized topological invariant presents non-zero quantized Hall conductance in the transport measurement \cite{15,16}.  
Besides, the BEC shows extreme robustness against the disorders \cite{Prodan_1,Prodan_2,loc_3,loc_4,loc_5,loc_7,TAI_1,TAI_2,TAI_theory,TAI_appli_1,TAI_appli_2,TAI_appli_3,mobility_gap,quasi_1,quasi_2,quasi_3}. 

Extending the topological band theory to nonlinear systems brings about exotic phenomena as well. 
Precisely, Refs.\cite{NL_wave_1,NL_wave_2,NL_wave_3,NL_wave_4,NL_wave_5,NL_wave_6,NL_wave_7,NL_wave_8,NL_wave_9} has recently studied the interaction between topology and nonlinearity of the eigenvectors, 
clarifying the occurrence of topological synchronization brought about by the interplay between nonlinearity and topology \cite{NL_wave_5}. 
In spite of extensive efforts have been made as described above, the interaction between the topology and nonlinear eigenvalues, 
which represent another form of nonlinearity \cite{Kuzmiak_Metal-PhC_PRB(1994),Huang_Mass-in-Mass_IJES(2009)}, has been seldom investigated. 
Very recently, Isobe et al. studied the BEC for the two-dimensional nonlinear eigenvalue systems with spin-orbit coupling \cite{Isobe}. 
The nonlinear BEC was established by introducing the auxiliary eigenvalues. The work provides a 
motivation for us to study whether there are nonlinear eigenvalues of BEC in two-dimensional systems beyond the spin-orbit 
coupling mechanism. We note that the Haldane model \cite{1} is topological system without spin-orbit coupling. Therefore, 
we will take the Haldane model as an medium to address the aforementioned issue.

This paper is organized as follows. Section \ref{S2} introduce the nonlinear eigenvalues problem of Haldane model. 
Section \ref{S3} presents the numerical results about edge states and phase diagram, and the analytical phase boundary. 
Section \ref{S4} presents our discussions and summary.

\begin{figure}[!htb]
\centering 
\subfigure[]{
\includegraphics[scale=0.4]{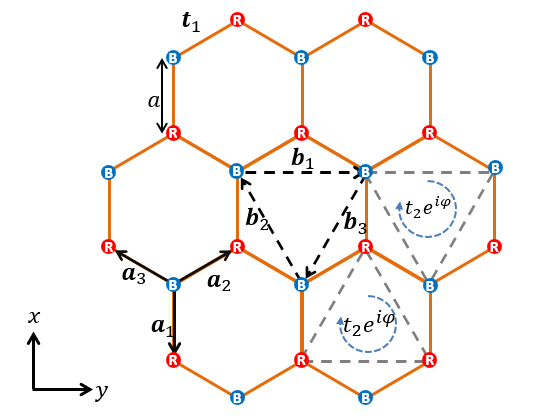} }
\subfigure[]{
\includegraphics[scale=0.4]{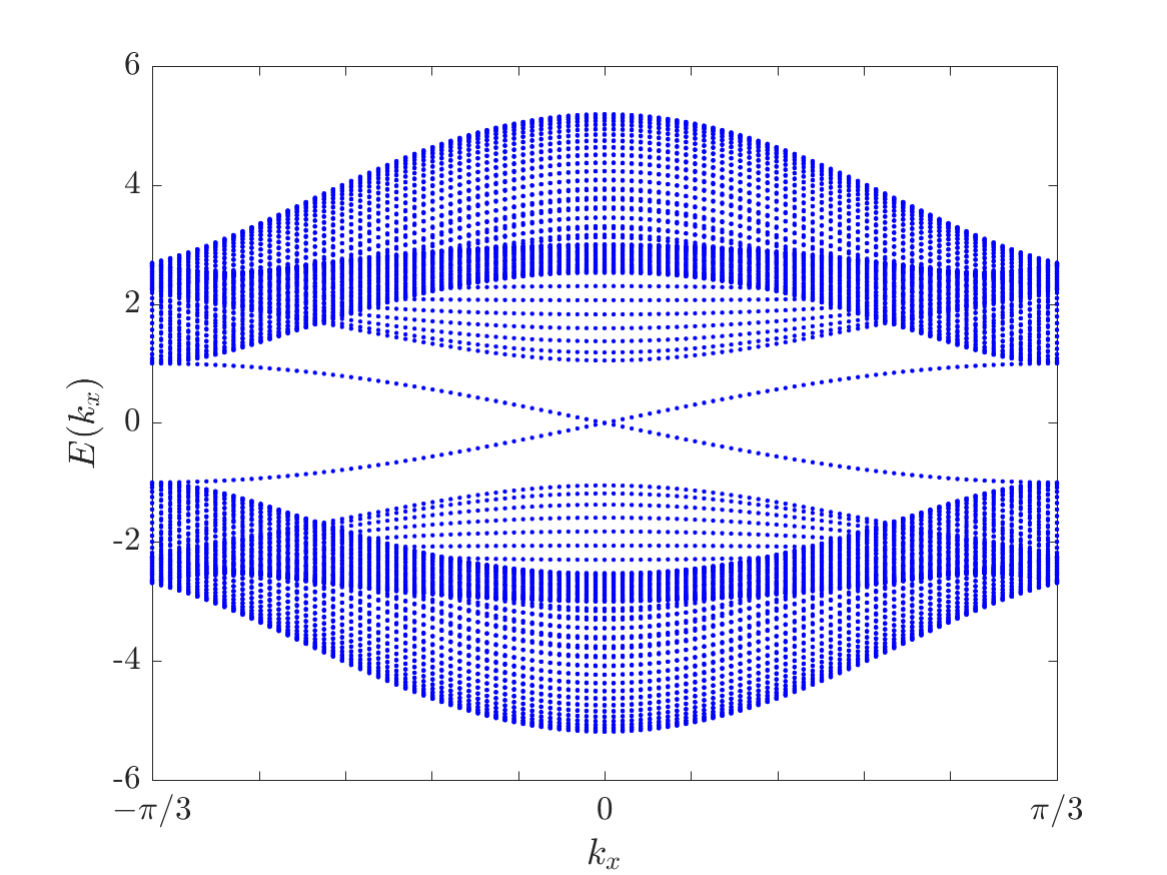}}
\caption{(a) Sketch of the Haldane lattice. ${\rm A}$ and ${\rm B}$ are two types of sublattice sites. 
$\bm{a_{n=1,2,3}}$ and $\bm{b_{n=1,2,3}}$ are lattice vectors. The bond length is set as $a=1$. $t_{1}$ is the hopping strength between 
nearest-neighbor sites (set as the unit of energy), and $t_{2}e^{-i\varphi}$ is the hopping strength between two same type sites. 
The on-site potential at ${\rm A}$ site is $M$ and the one at ${\rm B}$ site is $-M$.
(b) Armchair edge energy spectrum of the Haldane under $t_{2}=t_{1}$, $\varphi=\pi/2$, and $M=0$. 
}\label{f1}
\end{figure}

\section{nonlinear eigenvalues problem of Haldane model}\label{S2}
Here, we take the same strategy as told in Ref. \cite{Isobe} to analyze the nonlinear BEC of the Haldane model. 
Similarly, we discuss the BEC between the gapless edge states and the auxiliary topological bands by introducing 
the auxiliary eigenvalues. The nonlinear eigenvalues problem of the Haldane model is established by the 
following nonlinear equation 
\begin{equation}\label{eq1}
H(\mathbf{k})\ket{\psi}=\omega S(\omega, ~\mathbf{k})\ket{\psi}, 
\end{equation}
where $H(\mathbf{k})$ is the Hamiltonian matrix of the Haldane model, $S(\omega,~\mathbf{k})$ 
is the overlap matrix, $\mathbf{k}$ is the momentum, $\ket{\psi}$ is the eigenvector, $\omega$ is the 
nonlinear parameter. $H(\mathbf{k})$ is given by 
\begin{equation}
H(\mathbf{k})=\left(
\begin{array}{cc}
d_3 & d_1-id_2 \\
d_1+id_2 & -d_3
\end{array}
\right)
\end{equation}
The Hamiltonian elements $d_{1}$, $d_{2}$, and $d_{3}$ are 
\begin{equation*}
\begin{aligned}
d_{1}&=  \sum_{n=1,2,3}t_{1}\cos(\mathbf{k}\cdot a_{n}), \\
d_{2}&=  \sum_{n=1,2,3}t_{1}\sin(\mathbf{k}\cdot a_{n}), \\
d_{3}&= M-\sum_{n=1,2,3}2t_{2}\sin(\varphi)\sin(\mathbf{k}\cdot b_{n}),
\end{aligned}
\end{equation*}
where $t_{1}$ is the unit of energy and  
\begin{equation}
\begin{aligned}
& a_{1}=\left(
\begin{array}{cc}
0 \\
-1 
\end{array}
\right), ~
a_{2}=\left(
\begin{array}{cc}
\frac{\sqrt{3}}{2} \\
\frac{1}{2}
\end{array}
\right),~
a_{3}=\left(
\begin{array}{cc}
-\frac{\sqrt{3}}{2} \\
\frac{1}{2}
\end{array}
\right), \\
& b_{1}=\left(
\begin{array}{cc}
\sqrt{3} \\
0 
\end{array}
\right), ~
b_{2}=\left(
\begin{array}{cc}
-\frac{\sqrt{3}}{2} \\
\frac{3}{2}
\end{array}
\right),~
b_{3}=\left(
\begin{array}{cc}
-\frac{\sqrt{3}}{2} \\
-\frac{3}{2}
\end{array}
\right),
\end{aligned}
\end{equation}
are lattice vectors, and $t_{2}e^{i\varphi}$ is the hopping strength $t_{2}e^{-i\varphi}$ is the hopping strength 
between two same type sites and $M$ is the strength of on-site potential.

In fact, solving Eq.~(\ref{eq1}) is equivalent to solving $P(\omega,~\mathbf{k})\ket{\psi}=0$, 
where $P(\omega,~\mathbf{k})=H(\mathbf{k})-\omega S(\omega,~\mathbf{k})$. To analyze the 
BEC of Eq.~(\ref{eq1}), it is helpful to introduce the auxiliary eigenvalues $\lambda$, and the 
nonlinear eigenvalue problem becomes 
\begin{equation}\label{eq2}
P(\omega,~\mathbf{k})\ket{\psi}=\lambda\ket{\psi}. 
\end{equation}  
We shall remember that $\lambda$ is an auxiliary quantity that only has physical meaning at $\lambda=0$. Therefore, 
the above eigenvalue problem is transformed into finding the solution of $\lambda=0$. 

In Ref.~\cite{Isobe}, in establishing the above analysis process, the authors made an existence assumption of $\lambda=0$, 
such that one can observe the emergence and disappearance of gapless edge states at $\lambda=0$. 
For the nonlinear eigenvalue problem of Haldane model, we argue that the assumption of $\lambda=0$ is valid as well. 
At first, the energy spectrum of the Haldane model presents inversion symmetry. We consider a grip 
geometry of the Haldane lattice (see sketch in Fig.~\ref{f1}(a)), leaving periodic boundary condition in the $x$ direction and 
open boundary condition in the $y$ direction (armchair edge). Hence, the 
lattice constant is $3a$. The armchair edge spectrum $E(k_{x})$ as a function of the 
momentum $k_{x}$ is plotted in FIg.~\ref{f1}(b). As seen that the spectrum is symmetric with respect to $E(k_{x})=0$, presenting the 
inversion symmetry, and the edge states inevitably cross $E(k_{x})=0$. Secondly, we choose an overlap matrix $S$
only depending on the nonlinear parameter $\omega$, which is given by 
\begin{equation}\label{eq3}
S(\omega)=\left(
\begin{array}{cc}
1-M_{s}(\omega) & 0 \\
0 & 1+M_{s}(\omega)
\end{array}
\right),
\end{equation}
where $M_{s}(\omega)=M_{1}\tanh(\omega)/\omega$. From the expression of $\omega\pm \omega M_{s}(\omega)$, we know that 
they are monotonic with the change of $\omega$. When the nonlinearity is weak, i.e., 
$\omega$ is small, the eigenvalues of $S(\omega)$ are slow-varying with respect to $\omega$ (see ). Therefore, 
the up and down translation of the $\lambda$ spectrum caused by the nonlinear effect is small, 
and  we can observe the physical edge states at $\lambda=0$. In addition, the choice of $S(\omega)$ has been proved 
to be feasible to establish the nonlinear BEC in Ref. \cite{Isobe}. In the following, without loss of generality, we fix 
the parameters $t_{2}=M_{1}\equiv t_{1}$ and $\varphi=\pi/2$ to analyze the nonlinear BEC of the Haldane model.

\begin{figure}[!htb]
\centering 
\includegraphics[width=0.5\textwidth]{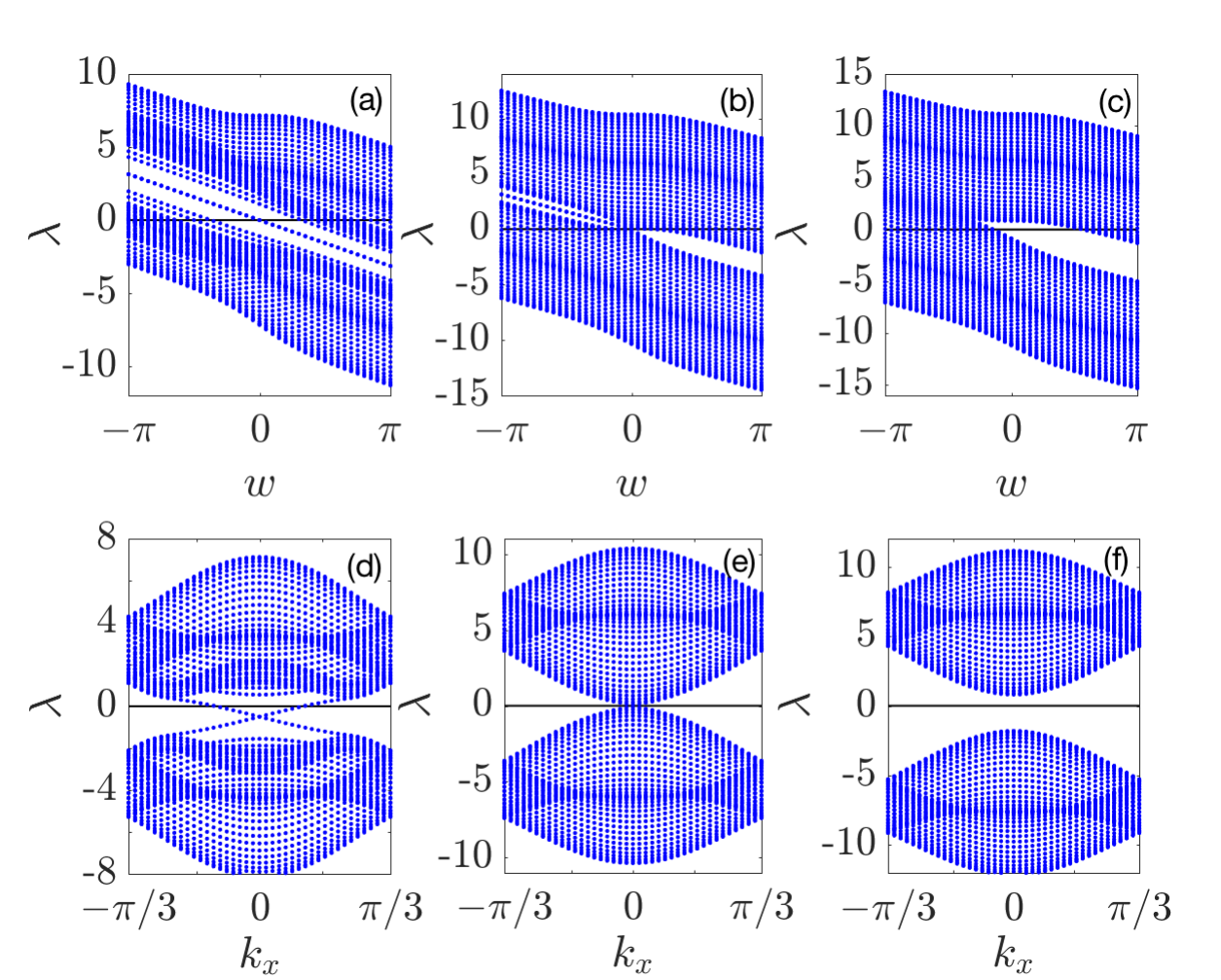}
\caption{Auxiliary $\lambda$ spectrum of the Haldane model, shown with blue dots. The horizontal black lines 
are reference lines at $\lambda=0$. (a)-(c): Presenting $\lambda$-$\omega$ spectra under $k_{x}=0$ with 
$M=2t_{1}$, $M=3\sqrt{3}t_{1}$, and $M=6t_{1}$, respectively. (d)-(f): Presenting $\lambda$-$k_{x}$ spectra 
under $\omega=0.5$, $\omega=0$, and $\omega=0.5$, respectively. 
}\label{f2}
\end{figure}

\section{nonlinear bulk-edge correspondence}\label{S3}
We start by analyzing the auxiliary $\lambda$ spectrum under different $M$. Similarly, the $\lambda$ spectra 
in the following are plotted by selecting a strip geometry with armchair edge in $y$ direction.  
To plot $\lambda$-$\omega$ spectra, we choose $k_{x}=0$ as an example. Under $M=2t_{1}$, 
the corresponding $\lambda$ spectrum as a function of the nonlinear parameter $\omega$ is plotted 
in Fig.~\ref{f2}(a). As seen that there are edge states crossing $\lambda=0$ under moderate values of $\omega$. 
To see the edge state clearly, we plot the $\lambda$ as a function of $k_{x}$ under $M=2t_{1}$ and $\omega=0.5$ 
in Fig.~\ref{f2}(d). Intuitively,  $\lambda=0$ is within the bulk gap in the $\lambda$ spectrum and there are a pair of edge states at $\lambda=0$, 
presenting non-trivial topological property. With $M=3\sqrt{3}t_{2}$, the $\lambda$ spectrum versus $\omega$ is presented 
in FIg.~\ref{f2}(b). We can see that the bands 
of $\lambda=0$ close at $\omega=0$. Equivalently, the bands in the $\lambda$ spectrum touch 
at $\lambda=0$ and $k_x=0$ (see Fig.~\ref{f2}(e) for details). For larger $M$, such as $M=6t_1$, we plot the 
the $\lambda$ spectrum with respect to $\omega$ in Fig.~\ref{f2}(c). It shows that there is no any edge state, although $\lambda=0$ 
is within a distinct gap of the $\lambda$ spectrum, presenting trivial topological property. Similarly, we can see the feature in 
the $\lambda$-$k_{x}$ spectrum as well. As Fig.~\ref{f2}(f) shows, $\lambda=0$ is within the band gap of the auxiliary $\lambda$ spectrum, 
but no any edge state crosses it.

\begin{figure}[!htb]
\centering 
\includegraphics[width=0.5\textwidth]{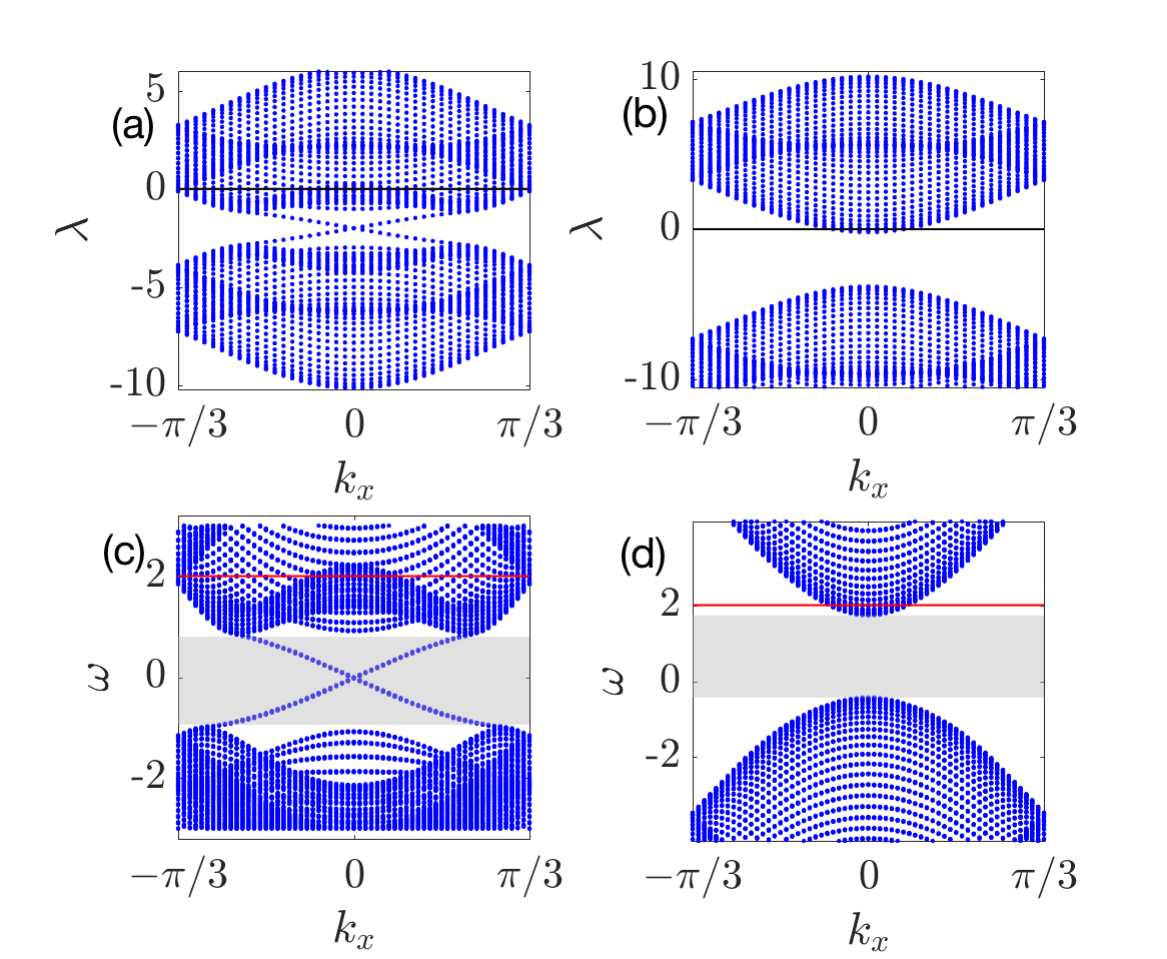}
\caption{(a) and (b): Armchair edge $\lambda$ spectra under $\omega=2$ with $M=2t_{1}$ in (a) and $M=6t_{1}$ in (b), respectively. The 
horizontal black lines are the $\lambda=0$ reference lines.  
(c) and (d): Band structures of $\omega$ versus $k_{x}$ extracted from $\lambda=0$ with $M=2t_{1}$ in (a) and $M=6t_{1}$ in (b), respectively. The gray regions 
show the band gaps. The horizontal red lines are the $\omega=2$ reference lines. 
}\label{f3}
\end{figure}

According to conventional principle of bulk-edge correspondence \cite{13,14}, the emergence of 
edge states can be forecasted by the energy band Chern number, and the magnitude of the Chern number counts the 
number of the paired edge states. Next, we check the correspondence between the Chern number of the bulk band 
of $\lambda$ and the emergent edge states at $\lambda=0$. The Chern number of the band in the auxiliary $\lambda$ spectrum 
below $\lambda=0$, namely $C_{1}(\omega)$, is defined as 
\begin{equation}
C_{1}(\omega)=\frac{1}{2\pi} \oint_{{\rm \partial}_ {\it 1BZ}} {\bm A}_{1}(\mathbf{k})~d\mathbf{k}, 
\end{equation}
where ${\rm \partial}_{\it 1BZ}$ means the boundary of the first Brillouin zone, and 
${\bm A}_{1}=-i\langle \psi_{1}(\mathbf{k})|\nabla_{\mathbf{k}}|\psi_{1}(\mathbf{k})\rangle$ with $\ket{\psi_{1}(\mathbf{k})}$
being the corresponding eigenvector. We analytically and numerically calculate $C_{1}(\omega=2)$ of the band below $\lambda=0$ in 
Fig.~\ref{f2}(d) and Fig.~\ref{f2}(f), and find $C_{1}=1$ and $C_{1}=0$, respectively. Here the analytical $C_{1}$ can be 
available by the singularity expansion method \cite{CN_cal_1,CN_cal_2}. It means that the correspondence 
between the number of paired edge states and the Chern number of the band below $\lambda=0$ is valid in the 
nonlinear eigenvalue problem of Haldane model.

What we have discussed before are the cases where $\omega$ are relatively small, and we find that when $\omega$ is relatively large, 
the system will enter the metallic phase, which can not be characterized by the band Chern number. Taking $\omega=2$ and $M=2t_{1}$ 
as an example, we plot the $\lambda$ spectrum as a function of $k_{x}$ under armchair edge in Fig.~\ref{f3}(a). As can be seen that the 
states at $\lambda=0$ has been embed into the bulk of the system (see the 
horizontal black reference lines). We name this state is the metallic phase. The metallic phase appears in $C_{1}=0$ case as well. Considering 
$\omega=2$ and $M=6t_{1}$, we plot the corresponding $\lambda$ spectrum in Fig.~\ref{f3}(b). It is seen that the state at $\lambda=0$ are 
embed into the bulk of system. Similarly, this metallic characteristic can be reflected from the $\omega$-$k_{x}$ spectrum as well. 
Figure \ref{f3}(c) presents the band structure of $\omega$ as the varying of $k_{x}$ under $M=2t_{1}$. The data are extracted from $\lambda=0$. 
As it shows, the edge states only exist in a small regions of the nonlinear parameter $\omega$ (the gray region shows) where there are bulk energy gaps, 
and the presence of the edge state can be interpreted by $C_{1}=1$. It means that there exists nonlinear BEC between the $\omega$-$k_{x}$ spectrum and 
the energy band Chern number as well.  For strong nonlinearity, 
there is no any edge state but bulk states (see the $\omega=2$ horizontal red line for example), showing the metallic feature of the system. 
In addition, similar nonlinear BEC and metallic characteristic can be seen in the $M=6t_{1}$ case as well. We plot the corresponding band structure 
of $\omega$ as a function of $k_{x}$ in Fig.~\ref{f3}(d). The data are still extracted from $\lambda=0$. Intuitively, there exists nonlinear BEC. 
When $\omega$ is small, there is a bulk energy gap (see the gray region), and the absence of edge state can be interpreted by $C_{1}=0$.  
For strong nonlinearity, such $\omega=2$ (the horizontal red reference line), the corresponding states are embed in the bulk of the system, 
presenting the metallic property. 

\begin{figure}[!htb]
\centering 
\includegraphics[width=0.5\textwidth]{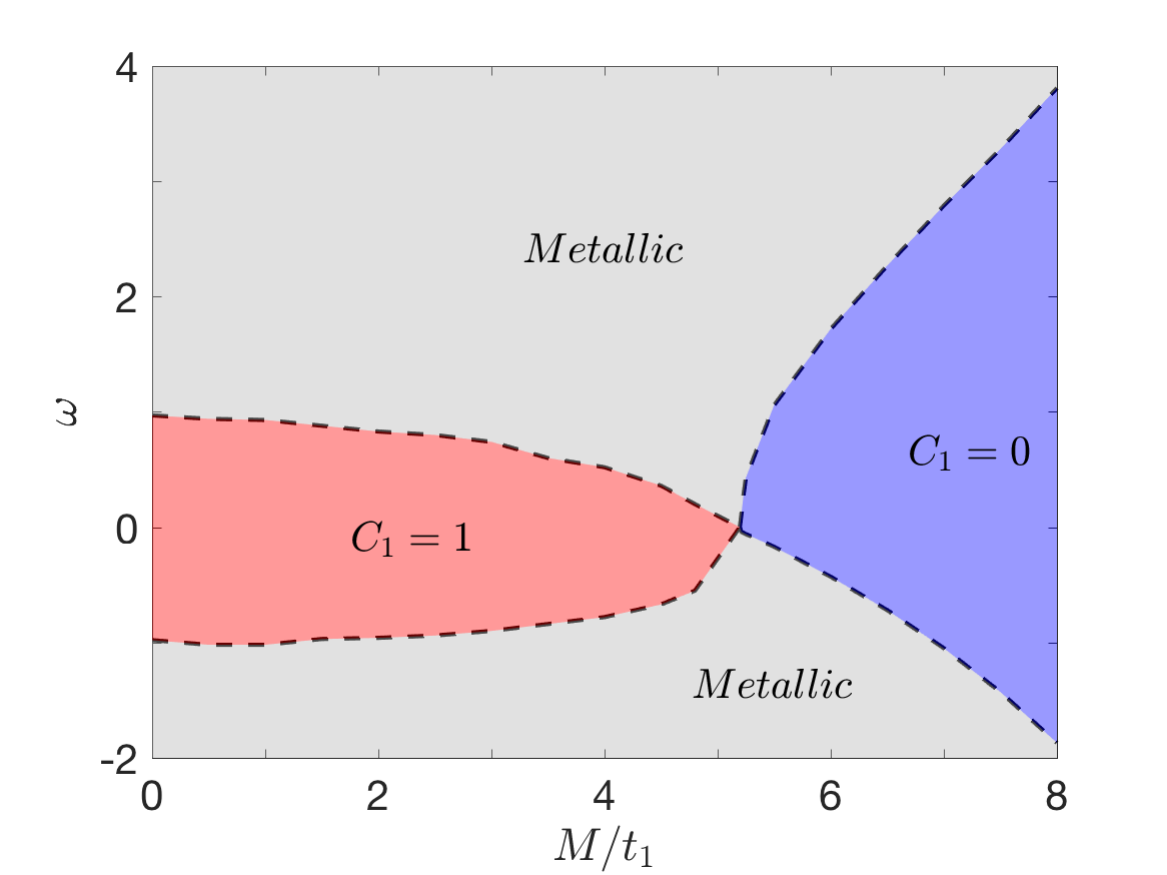}
\caption{ Phase diagram of the nonlinear Haldane model. The red region denotes the topological non-trivial with $C_{1}=1$. 
The blue region denotes the topological trivial phase with $C_{1}=0$. The gray region  denote the metallic phase. 
}\label{f4}
\end{figure}

By analyzing the band structures of the $\lambda$ and $\omega$ spectra under more discrete parameter points, the phase of the 
nonlinear Haldane model is plotted in the $\omega$-$M$ parameter space, which is shown in Fig.~\ref{f4}. We determine that 
the nonlinear system contains three phases: the topological nontrivial phase with $C_{1}=1$ (red region), the topological trivial phase with $C_{1}=0$ (blue region),  
and the metallic phase (gray region). The black dashed lines are the phase boundaries between the metallic phase and the topological phases. 
From the phase diagram, we can intuitively see that the topological nontrivial phase is more sensitive to the nonlinearity compared to the 
topological trivial phase. The topological nontrivial phase only exists in the cases where the nonlinear parameters $\omega$ are less than one, 
while the trivial phase can survival in the cases where $\omega$ far larger than one. When the nonlinear parameter $\omega$ is fixed at 
a finite nonzero value, as the increase of the on-site potential strength, the system can undergo the transition from the $C_{1}=1$ phase to the 
metallic phase, and finally to the $C_{1}=0$ phase. It is also feasible to continuously tune the nonlinear parameters and the strength of the potential, 
achieving a direct transition from $C_{1}=1$to $C_{1}=0$ without experiencing the metallic phase. 

\section{Summary}\label{S4}
Herein, we have studied the nonlinear eigenvalue problem of the Haldane model. We find that there is nonlinear bulk-edge correspondence 
in the absence of spin-orbit coupling. When the nonlinearity is within the threshold, the emergence and disappearance of the edge states can be characterized 
by the Chern number of the auxiliary energy band. When nonlinearity exceeds the threshold, this nonlinear system will enter the metallic phase. 
Compared to the topological trivial phase, the topological non-trivial phase is more fragile to the nonlinearity, because it only appears in the cases 
where the nonlinearity is relative weak. Our work enriches the study of the bulk-edge correspondence of nonlinear eigenvalues of two-dimensional systems. 
Noting that the Haldane model has been experimentally realized \cite{Esslinger}, we expect that nonlinear body-edge correspondence of the Haldane model can 
be observed on similar experimental platforms in the near future.

We acknowledge support from NSFC under Grants No. 11835011, No. 12174346, and No. 51405449.

\bibliography{references.bib}
\end{document}